\documentclass[12pt] {article}

\newcommand{\beq}{\begin{equation}}
\newcommand{\feq}[1]{\label{#1} \end{equation}}
\newcommand{\beqr}{\begin{eqnarray}}
\newcommand{\feqr}{\end{eqnarray}}
\def\non{\nonumber}
\def\noi{\noindent}
\newcommand{\rf}[1]{~(\ref{#1})}
\def\pr{^{\prime}}
\def\np#1#2#3{Nucl. Phys. {\bf{B#1}} (#2) #3}
\def\cm#1#2#3{Comm. Math. Phys. {\bf{#1}}, (#2), #3}
\def\pr#1#2#3{Phys. Rep. {\bf{#1}}, (#2), #3}
\def\prev#1#2#3{Phys. Rev. {\bf{D#1}}, (#2), #3}
\def\prl#1#2#3{Phys. Rev. Lett.{\bf{#1}}, (#2), #3}
\begin{document}

\renewcommand{\thefootnote}{\fnsymbol{footnote}}

\begin{center}

{\LARGE \bf The gravitational chiral anomaly of spin-1/2 field in the 
presence of twisted boundary conditions for ordinary field theory.}\\[6mm]

\large{Agapitos Hatzinikitas}\footnote [2] 
{e-mail: ahatzini@cc.uoa.gr}\\[8mm]

{\it University of Athens, \\
Nuclear and Particle Physics Division\\
Panepistimioupoli GR-15771 Athens, Greece.}\\[8mm]

{\small \bf Abstract}\\[3mm]

\parbox{6.2in}
{\small We calculate the chiral anomaly in the neighbourhood of the fixed 
point space $\mathcal{M}_{h}$ which is constructed by the group action of a 
discrete symmetry $h$ on a compact manifold $\mathcal{M}$. The Feynman 
diagrams 
approach for the corresponding supersymmetric quantum mechanical system with 
twisted boundary conditions is used. The result we derive in this way agrees 
with the generalization of the ordinary index theorem (the G-index theorem) 
on the spin complex.}

\end{center}

%\newpage

\section{Introduction}

\renewcommand{\thefootnote}{\arabic{footnote}}
\setcounter{footnote}{0}

Anomalies of quantum field theories in the path integral approach can be
interpreted, according to Fujikawa \cite{fuji}, as a symptom of the impossibility of
defining an invariant measure in the functional integral which determines the graviton theory
under investigation. Therefore symmetries of the classical action cease to be
conserved at the quantum level. For a massless fermion $\psi$, in even-dimensions,
the action of the field theory is given by:
\beqr
\mathcal{L}=-(\det e^{\alpha}_{\mu}) \bar{\psi} e^{\mu}_{\alpha}(x) \gamma^{\alpha} \nabla_{\mu}\psi
\label{actfrm}
\feqr

\noi where $\nabla_{\mu} \psi= \partial_{\mu}\psi + 
\frac{1}{4} \omega_{\mu ab}(e) \gamma^a \gamma^b \psi$, with $\omega_{\mu ab}$ the
spin connection. Chiral anomaly is associated with the infinitesimal local transformation
$\psi \rightarrow \psi+ \alpha (x) \gamma_5 \psi$ where $\alpha$ is a complex space-time
 dependent function.  
Fujikawa found that for  a spin-$\frac{1}{2}$ loop the chiral anomaly is given by 
the regulated Jacobian $\lim_{\beta \rightarrow 0}Tr\left[\gamma_5 \left(-\frac{\beta}{\hbar} 
D\!\!\!\!\slash  D\!\!\!\!\slash \right) \right]$, and $D\!\!\!\!\slash D\!\!\!\!\slash$
is the regulator. 

The computation of anomalies using Fujikawa's scheme was soon realized to be very cumbersome 
when trying to evaluate traces in n-dimensions involving  products of $\gamma$ matrices. A new
procedure based on one-dimensional quantum mechanics was then proposed by Alvarez-Gaum\'{e} 
and Witten \cite{alvarez}. According to this method the operators $\gamma_5$,
 $\nabla_{\mu}$, $x^{\mu}$, $\gamma^{\mu}$ 
were represented by operators of a corresponding quantum mechanical model, and by 
turning these operator expressions into path integrals, one finds that anomalies of 
quantum field theories can be written in terms of Feynman diagrams for certain sigma 
models on the worldline.       

In the case of chiral anomalies the partition function receives contributions from
periodic bosonic coordinates and the projection operator $(-1)^F$ replaces
the antiperiodic boundary conditions for the fermions with periodic ones. It 
is worth noting that chiral anomalies due to their topological nature are 
insensitive to the method used for calculating them in contrast to the trace 
anomalies which are not.

In this article we introduce twisted boundary conditions for bosons and 
fermions induced by automorphic maps of the base manifold. For convenience
we consider $Z_N$ as the discrete isometry group. The twisted chiral anomaly 
will be given by the trace:
\beq
An(chiral)=\lim_{\beta \rightarrow 0} Tr \left((-1)^F \hat{\mathcal{P}} 
e^{-\frac{\beta}{\hbar}H} \right)
\feq{trace}    

\noi where $(-1)^F$ is the $4n$ analogue of $\gamma_{5}$, $F$ is the 
one-dimensional fermion number, $\mathcal{P}$ is a projection operator which 
selects those states
of the total Hilbert space that remain invariant under the group action and 
$H$ is the $N=1$ supersymmetric Hamiltonian. The
paths of the points that belong to the group invariant subspace dominate in 
the functional integral. Away from the fixed point space the corresponding
contributions are exponentially suppressed.

The extra contribution to the chiral anomaly stems from the vacuum expectation
value of a vertex that couples the complex zero modes of periodic fermions 
with the 
complex bosonic quantum fluctuations. Writing then the chiral anomaly as 
a double sum of the 
relevant twisted worldline graphs all satisfying the same boundary 
conditions, we recover the standard result of the literature 
\cite{eguchi,witten1, witten2}.

\section{The group action and the Hilbert space}

We consider a compact Riemannian manifold $\mathcal{M}$ (which might be 
curved with nontrivial metric) of even dimension 
$dim \mathcal{M}=4n$ with coordinates described by the real bosonic fields 
$x^{\mu}(\tau)$. In the supersymmetric quantum mechanics we also have 
Majorana fermions living on the sections of the pullback of the tangent bundle
of the tangent space $T\mathcal{M}$ with 
$dim T\mathcal{M}=4n$. When our base manifold has a discrete isometry $h\in
 G$
then there is a submanifold $\mathcal{M}_h$ (possibly of non vanishing 
dimension) that remains invariant under the corresponding group action. 
We may also assume that the manifold $\mathcal{M}$ is K\"{a}hler 
equipped with a complex structure $\mathcal{J}$ preserved under the action 
of $h$:
\beq
h_{\ast}\circ \mathcal{J}= \mathcal{J} \circ h_{\ast}
\feq{coms}

\noi where the asterisk represents push-forward of the vectors. If 
the normal bundle $\mathcal{N}_{\mathcal{M}} (\mathcal{M}_h)$ decomposes into
flat complex bundles then the real dimension at a point $p$ of 
$\mathcal{M}_h$ is:
\beq
dim \mathcal{M}_h |_{p}=m=4n-2s.
\feq{dim}

\noi The above constraint implies that $\mathcal{M}_h$ is of even dimension. 
The specific choice for the dimension of the base manifold is implied by the
existence of the untwisted chiral anomaly only in $4n$-dimensions. Note that
we may use for $\mathcal{M}_h$ the same $\mathcal{J}$ as that used for
$\mathcal{M}$.    

Let $h$ be an orientation preserving map which generates the $Z_{N}$ 
transformation realised by the rotation:
\beqr
\left( \begin{array}{c} 
x^{2i}(\tau) \\ x^{2i-1}(\tau) \end{array} \right) 
\rightarrow \left( \begin{array}{cc}
\cos \theta_{i} & -\sin \theta_{i} \\
\sin \theta_{i} & \cos \theta_{i}
\end{array} \right)
\left( \begin{array}{c}
x^{2i}(\tau) \\ x^{2i-1}(\tau) \end{array} \right) 
\label{tran}
\feqr

\noi where $i=1,\ldots ,dim \mathcal{N}_{\mathcal{M}}(\mathcal{M}_h)/2$, 
$\theta_{i}=\frac{2\pi}{N} k_i$, $k_{i}=0,\ldots ,N-1$ and $U_i$ satisfies 
$U_{i}^{N}=1$. Since $U_i$ is an orthogonal matrix it can be diagonalized 
by a unitary matrix $M_i$:
\beq
M_{i}U_{i}M^{\dagger}_{i}=U_{i,diag}
\feq{trand} 

\noi  from which one obtains:
\beqr
U_{i,diag}= \left( \begin{array}{ll}
g^{k_i} & 0   \\
0       & g^{-k_i} 
\end{array} \right)  
\label{diagm}
\feqr

\noi with $g=e^{2i\pi / N}$ \footnote{The explicit form of $M_i$ is:
$M_i=\left( \begin{array}{cc} \alpha & i\alpha
 \\ i\alpha^{\ast} & \alpha^{\ast} \end{array} \right)$ with the 
condition that $\|\alpha\|=\frac{1}{\sqrt{2}}$.}
. In this way the eigenvalues of $U_i$ fall into 
two subsets $\left\{g^{k_i} \right\}$ and $\left\{g^{-k_{i}} \right\}$.

Thus, adopting complex notation for the bosons on the normal bundle we impose
 the following twisted boundary conditions:
\beqr
\left( \begin{array}{c}
X^{i} \\ X^{\bar{i}} \end{array} \right)= \frac{1}{\sqrt{2}}
\left( \begin{array}{c} x^{2i}+ix^{2i-1} \\
x^{2i}-ix^{2i-1} \end{array} \right) \rightarrow  
\left( \begin{array}{ll}
g^{k_i} & 0   \\
0       & g^{-k_i} 
\end{array} \right)
\left( \begin{array}{c}
X^{i} \\ X^{\bar{i}} \end{array} \right)
\label{twb}
\feqr

\noi and similar twisted conditions hold for the fermions on the tangent 
space. Thus the base manifold coordinates split into the following
set:
\beq
x^{\mu}=(x^{l}, X^{i}, X^{\bar{i}})
\feq{cooba}

\noi where $x^l$, $l=1,\cdots,m$ describe the fixed point manifold. These
coordinates can be decomposed into background trajectories obeying the 
boundary condition $x^{l}_{bg}(\tau=-1)=x^{l}_{bg}(\tau=0)$ and a quantum
 fluctuating part with $q^{l}(\tau=-1)=q^{l}(\tau=0)=0$. On
the other hand the $X^{i}, X^{\bar{i}}$, $i,\bar{i}=1,\cdots,(4n-m)/2$ 
describe the manifold normal to $\mathcal{M}_g$ and the appearance of 
twisted boundary conditions does not allow the existence of a classical 
part but permits only a quantum part (denoted by $Q^{i}, Q^{\bar{i}}$ 
respectively).

Recalling the standard lore or orbifold theories \cite{dixon}, each 
$k_i$-sector (one for each conjugacy class in the group) has its own mode 
expansion and 
Hilbert space $\mathcal{H}_{k_i}$ \footnote{In the non-Abelian case the 
twisted sectors should be labeled by the conjugacy classes $\mathcal{C}_{i}$
of the group G. The projection operator is replaced by
$\hat{\mathcal{P}}=\sum_{i}\frac{1}{|N_{i}|} \sum_{g\in N_{i}} 
g\zeta_{\{h\}}$ where the centralizer $N_{i}$ is defined by
$N_{i}=\{g|h \in C_{i} : [g,h]=0 \}$ and the action of any $g\in N_{i}$
on $\zeta_{\{h\}}$ is defined by $g\zeta_{\{h\}} = (g\zeta_{ \{h\} },
l_{1}g l^{-1}_{1} \zeta_{ \{l_{1}h l^{-1}_{1}\} },\cdots)$.}. 
The total Hilbert space is the direct sum of $\mathcal{H}_{k_i}$:
\beq
\hat{\mathcal{H}^i}_{tot}=\oplus_{k_i=1}^{N}\mathcal{H}_{k_i}.
\feq{tohi}

\noi The physical Hilbert space $\mathcal{H}^{i}_{phys}$ consists of states 
that remain invariant under the action of all $\hat{h}_{k_i}$:
\beq
\hat{\mathcal{H}}^{i}_{phys}=\hat{\mathcal{P}}^i \mathcal{H}^i
\feq{phyhil}

\noi with the projection operator defined by:
\beq
\hat{\mathcal{P}}^i=\frac{1}{N}\sum_{k_i=1}^{N} \hat{h}_{k_i}.
\feq{proj}

\noi Notice that $\mathcal{P}^i$ is indeed a projection operator since
$(\hat{\mathcal{P}}^i)^2=\hat{\mathcal{P}}^i$ and $\hat{h}_{k_j}
\hat{\mathcal{P}}^i=\hat{\mathcal{P}}^i$.

\section{Interactions and propagators}

In the path integral approach and applying the background field formalism 
\footnote{In principle one could compute an infinite set of Feynman diagrams 
with a spin-$\frac{1}{2}$ field circulating in a loop and external gravitons 
couple at the vertices of the fermion loop. The
advantage of the background formalism is encaptulated in the fact that it takes 
the contributions of all vertices at once.} to the untwisted chiral anomaly 
one obtains \cite{kostas}:
\beqr
An(chiral)=\left(\frac{-i}{2\pi}\right)^{n/2} \int_{-1}^{0}\prod_{i=1}^{n} dx^{i}_{0}
\sqrt{g(x_{0})} \prod_{\alpha=1}^{n} d\psi^{\alpha}_{1,bg} 
<e^{-\frac{1}{2\beta\hbar}S_{int}}>
\label{anutc}
\feqr

\noi where $S_{int}=\int_{-1}^{0} \mathcal{L}_{int}(\tau) d\tau$ and
\beqr
\mathcal{L}_{int}(\tau) &=& \!\!\!\! \left[g_{\mu \nu}(x_0 +q)-g_{\mu \nu}(x_0)\right]
\left(\dot{q}^{\mu}(\tau) \dot{q}^{\nu}(\tau) + b^{\mu}(\tau) c^{\nu}(\tau) 
+ \alpha^{\mu}(\tau) \alpha^{\nu}(\tau)\right) \non \\
&+& \!\!\!\! \dot{q}^{\mu}(\tau) \omega_{\mu \alpha \beta}(x_0 +q) \psi^{\alpha}_{1,bg} 
\psi^{\beta}_{1,bg} \non \\
&-& \!\!\!\! \left(\frac{\beta \hbar}{2}\right)^{2} \!\!\! g^{\mu \nu}(x_0 +q)
\left(\Gamma_{\mu \rho}{}^{\sigma}(x_0 +q) \Gamma_{\nu \sigma}{}^{\rho}(x_0 +q) -
\frac{1}{2} \omega_{\mu}{}^{\alpha \beta}(x_0 +q) \omega_{\nu \alpha \beta}(x_0 +q)\right).
\label{sintu}
\feqr

\noi The expectation value $<>$ means that all quantum fields must be contracted using 
the appropriate propagators. In the expression\rf{sintu} $(b^{\mu}, c^{\nu})$ and $\alpha^{\mu}$ is
a set of anticommuting and commuting Lee-Yang ghosts respectively \footnote{The significance 
of such
ghosts is implied by the integration over the momenta $p_{\mu}(\tau)$ in the transition
from phase space to configuration space. A factor $\sqrt{g}$ is then produced  
and by exponentiating it, introducing the Lee-Yang ghosts, we are led to a perfectly regular term 
in the action.}. When we take the limit $\beta \rightarrow 0$ only one-loop graphs survive 
and the corresponding interaction is:
\beqr
\mathcal{L}_{int}(\tau)=\frac{1}{2}q^{\mu}(\tau) \dot{q}^{\nu}(\tau) R_{\mu \nu  ab}(\omega(x_0)) 
\psi^{a}_{1,bg} \psi^{b}_{1,bg}
\label{surac}
\feqr

\noi where $R_{\mu \nu  ab}(\omega(x_0))=2\partial_{[\mu}\omega_{\nu] ab}(x_0)$ in a frame with
$\omega_{\mu ab}(x_0)=0$ 

In our case the total interaction Langrangian density decomposes into an untwisted  
and a twisted part:
\beqr
\mathcal{L}_{int} &=& \mathcal{L}_{U}+\mathcal{L}_{T} \non \\
&=& \frac{1}{4\beta}\left(\psi^{a}_{1,bg} \psi^{b}_{1,bg}R_{ablk}(\omega) 
q^{l} \dot{q}^{k}
+ \frac{1}{2}\Psi^{i}_{1,bg} \Psi^{\bar{j}}_{1,bg} R_{i\bar{j}m\bar{n}} 
(\omega) Q^{m} \dot{Q}^{\bar{n}}   \right).
\label{intac}
\feqr

\noi The first term corresponds to the familiar interaction\rf{surac} on the base manifold 
while the second describes the interaction on the normal bundle.  

The propagator for bosons that move along the fibres $R^{2}$ of the normal 
bundle is:
\beqr
<Q^i (\tau) Q^{\bar{j}} (\sigma)>_{k_i}=-\beta \hbar g^{i\bar{j}}
\Delta_{k_i}^{b}(\tau-\sigma)
\label{bprop}
\feqr

\noi where $g^{i\bar{j}}$ is the metric on the normal bundle and 
$\Delta_{k_i}^{b}$ is the Green's function given by:
\beq
\Delta_{k_i}^{b}(\tau - \sigma)=\sum_m \frac{\phi_{m}(\tau) \phi^{\ast}_{m} 
(\sigma)}{\lambda_m}.
\feq{bgren}

\noi The $\lambda_m$'s are the eigenvalues of the kinetic operator 
$d^2 /d\tau^2$, $\Delta_{k_i}^{b}$ satisfies: 
\beq
\frac{\partial^{2}}{\partial \tau^{2}} \Delta_{k_i}^{b}(\tau - \sigma)
=\delta(\tau - \sigma)
\feq{eqd}

\noi and in addition, it is subjected to the boundary condition:
\beqr
\Delta_{k_i}^{b}(\tau,-1)=e^{i\theta_{i}} \Delta_{k_i}^{b}(\tau, 0).
\label{scon}
\feqr

\noi on the time interval $[-1,0]$. One then finds the solution:
\beq
\Delta_{k_i}^{b}(\tau-\sigma)=\sum_{n=-\infty}^{+\infty} \frac
{e^{i(2\pi n + \theta_i)(\tau-\sigma)}}{[i(2\pi n+\theta_i)]^2}
\feq{solb}

\noi from which by setting $k_i=0$ we recover the center-of-mass 
Green's function \cite{takis}:
\beqr
\Delta_{cm}^{b}(\tau-\sigma)=-\sum_{n=-\infty}^{+\infty}{^{\prime}} \frac
{e^{2i\pi n(\tau-\sigma)}}{(2\pi n)^2}=\frac{1}{2}(\tau-\sigma)
\epsilon(\tau-\sigma)-\frac{1}{2}(\tau-\sigma)^2 -\frac{1}{12}.
\label{comg}
\feqr

The fermionic propagator is:
\beq
<\Psi^i(\tau) \Psi^{\bar{j}}(\sigma)>_{k_i}=g^{i\bar{j}} \Delta_{k_i}^{f}
(\tau - \sigma)
\feq{ferpro}

\noi and $\Delta_{k_i}^{f}$ is given by:
\beq
\Delta_{k_i}^{f}(\tau-\sigma)=\sum_{n=-\infty}^{+\infty}
\frac{e^{i(2\pi n + \theta_i)(\tau-\sigma)}}{i(2\pi n + \theta_i)}.
\feq{frgr}

\section{Twisted chiral anomaly}

The untwisted chiral anomaly, in which both bosonic and fermionic fields are periodic,
can be written equivalently as in \cite{takis}:
\beqr
An(chiral) &=& \left(\frac{-i}{2\pi}\right)^{n/2} \int_{-1}^{0}\prod_{i=1}^{n} dx^{i}_{0}
\sqrt{g(x_{0})} \prod_{\alpha=1}^{n} d\psi^{\alpha}_{1,bg} 
\exp \left[ \frac{1}{2} Tr ln \left(\frac{R_{\mu \nu}/4}{\sinh(R_{\mu \nu}/4)} \right)
\right].
\label{chiue}
\feqr

\noi At first sight the expression\rf{chiue}
is obscured since for example $R^4$ could equally mean $tr(R^4)$ or 
$tr(R^2) tr(R^2)$. The recipee one has to apply is to write down the 
series for the logarithm, replace $R^m \rightarrow tr(R^m)$ everywhere, 
and only then one should take the exponential.

Using the interaction Lagrangian density\rf{intac} and the propagators\rf{bprop}, \rf{ferpro} 
one evaluates the total integrated chiral anomaly of the relevant worldline graphs to be 
proportional to:
\beqr
<e^{-\frac{1}{\hbar} \mathcal{S}_{int}}> &=& e^{\frac{1}{2}Tr ln \left(
\frac{R_{ij}/4}{\sinh (R_{ij}/4)}\right)}  \non \\
&\times& \exp \left[\frac{1}{N}
\sum_{k_{i}=1}^{N-1}\sum_{l=1}^{\infty} \left(-\frac{1}{\beta \hbar} 
\right)^{l}
\frac{(l-1)!}{l!} 2^{l-1} tr\left(\frac{R_{m\bar{n}}}{8} \right)^{l}
(-\beta \hbar)^{l} I_{k_i, l}\right]
\label{inact}
\feqr

where
\beq
I_{k_i, l}=\int_{-1}^{0} d\tau_1 \int_{-1}^{0} d\tau_2
 \cdots \int_{-1}^{0} d\tau_l \,
\Delta^{\bullet}_{k_i}(\tau_1 - \tau_2) 
\Delta^{\bullet}_{k_i}(\tau_2 - \tau_3)
\cdots \Delta^{\bullet}_{k_i}(\tau_l -\tau_1)
\feq{chan}

\noi and all bosons are twisted in the same way with 
$\Delta^{\bullet}_{k_i}(\tau-\sigma)=\frac{\partial}{\partial \sigma}
\Delta_{k_i}(\tau - \sigma)$. The combinatorial factors $(l-1)!$ and 
$2^{l-1}$ stand for the ways one can contract $l$ vertices and two q fields 
at $l$ vertices respectively. The multiple integral can be evaluated using
successively:
\beq
\int_{-1}^{0} \Delta^{\bullet}_{k_i}(\tau_1 -\tau_2)
\Delta^{\bullet}_{k_i}(\tau_2 -\tau_3) \, d\tau_2 =
\Delta_{k_i}(\tau_1 -\tau_3),
\feq{mul1}
\beq
\int_{-1}^{0} \Delta_{k_i}(\tau_1 -\tau_3) \Delta^{\bullet}_{k_i}
(\tau_3 -\tau_4) \, d\tau_3= \left(\frac{\partial}{\partial \tau_4} 
\right)^{-1} \Delta^{\bullet}_{k_i}(\tau_1 - \tau_4)
\feq{mul2}
\beqr
\vdots \non \\
\non
\label{vtel}
\feqr

\noi and the final integral reads:
\beqr
I_{k_i,l} &=& \int_{-1}^{0} \int_{-1}^{0} d\tau_1 d\tau_l 
\left(\frac{\partial}{\partial \tau_l} \right)^{-(l-3)}
\Delta_{k_i}(\tau_1 -\tau_l) \Delta^{\bullet}_{k_i}(\tau_l -\tau_1) \non \\
&=& \frac{(-1)^l}{(2i\pi)^l} \sum_{n=-\infty}^{+\infty} \frac{1}{(n+\alpha
_{k_i})^l}
\label{mull}
\feqr

\noi where $\alpha_{k_i}=\frac{k_i}{N}$. If we compare this result with the 
one derived from the untwisted sector \cite{takis} we observe that 
$I_{k_i,l}$ is no longer zero for odd values of $l$. To demonstrate this 
consider the case $I_{k_i,1}$ then:
\beqr
I_{k_i,1} &=& \int_{-1}^{0} \Delta^{\bullet}_{k_i}(0) d\tau_1=
\frac{1}{2i\pi}\sum_{n=-\infty}^{+\infty}\frac{1}{(n+\alpha_{k_i})} \non \\
&=& \frac{1}{2i\pi} \left[ f(1-\alpha_{k_i})-f(\alpha_{k_i}) \right]
=\frac{1}{2i} \cot(\pi \alpha_{k_i})
\label{i1}
\feqr

\noi where $f(z)=\frac{\Gamma ^{\prime}(z)}{\Gamma (z)}$ and $\Gamma(z)$
is the well known gamma function (see Appendix for details). Consider
now a closed q-loop with an odd number of vertices ($-\frac{1}{4\beta}
R_{\mu \nu a b}(x_0) \psi^{a}_{1,bg} \psi^{b}_{1,bg}= -\frac{1}{4\beta}
R_{\mu \nu}(x_0)$)
and perform all the contractions of the related quantum bosonic fields
(two at each vertex namely $q^{\mu}(\tau)$, $\dot{q}^{\nu}(\tau)$).
There always be one vertex in the end for which upon contracting the attached bosonic
fields (at a given time) will give, apart from a numerical factor, a result 
$\propto g^{\mu \nu}R_{\mu \nu} R^{2k} =0$. Only parity violating amplitudes are anomalous and 
thus the twisted chiral anomaly receives 
contributions exclusively from even values of $l$. Therefore the anomaly is propotional to:
\beqr
<A_{twisted}> &=& \frac{1}{N} \exp \left[ \sum_{k_i=1}^{N-1}
\sum_{l=1}^{+\infty}
\left(\frac{1}{4} \right)^{2l} 2^{2l-1} \frac{(2l-1)!}{(2l)!} 
Tr\left(\frac{R_{m\bar{n}}}{2}\right)^{2l} I_{k_i,2l} \right] \non \\
&=& \exp \left[ \frac{1}{2} \ln (\frac{1}{N^2}) +
\frac{1}{2} Tr \left( \sum_{k_i=1}^{N-1} \sum_{l=1}^{+\infty} 
\frac{z^{2l}}{2l} \sum_{n=-\infty}^{+\infty} \frac{1}{(n+\alpha_{k_i})^{2l}}
\right) \right]
\label{twch} 
\feqr

\noi where $z=\frac{R_{m\bar{n}}}{8i\pi}$. Performing first the summation over 
l, then over n and making use of the identity (see Appendix for the proof):
\beqr
\left(1\pm \frac{z}{\alpha_i} \right) \prod_{n=1}^{+\infty}
\left(1\pm \frac{z}{n+\alpha_{k_i}}\right)\left(1\mp \frac{z}{n -\alpha_{k_i}}
\right)=\frac{\sin(\pi(z\pm \alpha_{k_i}))}{\sin(\pi \alpha_{k_i})}
\label{baid}
\feqr

\noi yields:
\beqr
<A_{twisted}> &=& \exp \left[
\frac{1}{2} Tr \ln \frac{1}{N^2}\prod_{k_i=1}^{N-1} \left( 
\frac{\sin (\pi \tilde{\alpha}_{k_i})}
{i\sin (\frac{\tilde{\alpha}_{k_i}}{2})}
\frac{\sin (\pi {\tilde{\alpha}_{k_i}}^{\ast})}
{i\sin (\frac{\tilde{\alpha}_{k_i}^{\ast}}{2} )} \right) \right] \non \\
&=& \exp \left[ \frac{1}{2} Tr \ln \prod_{k_i=1}^{N-1} \left(
\frac{1}{2i\sin (\frac{\tilde{\alpha}_{k_i}}{2})} 
\frac{1}{2i\sin (\frac{{\tilde{\alpha}_{k_i}}^{\ast}}{2})} \right) \right] 
\non \\
&=& \exp \left[ Tr \ln \prod_{k_i=1}^{N-1} \left(
\frac{1}{2i\sin (\frac{\tilde{\alpha}_{k_i}}{2})} \right) \right] 
\label{fch}
\feqr

\noi where $\tilde{\alpha}_{k_i}=2\pi \alpha_{k_i} + \frac{iR_{m\bar{n}}}{4}$, 
 $\tilde{\alpha}_{k_i}^{\ast}$ is the complex conjugate of 
$\tilde{\alpha}_{k_i}$ and 
\beq
\prod_{k_i=1}^{N-1} \sin (\pi \alpha_{k_i})=\frac{N}{2^{N-1}}. 
\feq{prof}
             
\noi The last equality in \rf{fch} can be justified from the fact that:
\beqr
\sin (\pi \alpha_{k_i} + z) = \left \{ \begin{array}{ll}
\sin (\pi \alpha_{k_i +2}  -z) & \mbox{for $k_i +2 <k_i$} \\
\sin (\pi \alpha_{k_i}  -z)    & \mbox{for $k_i +2 =N$} \\ 
\sin (\pi \alpha_{k_i -2}  -z) & \mbox{for $k_i +2 >k_i$} 
\end{array} \right. .
\label{sines}
\feqr

\noi This expression is in agreement with the one found by Eguchi
 \cite{eguchi} and Witten \cite{witten1,witten2}.

\section{Conclusions}

We have extended the Feynman diagrams formalism developed by the authors
 \cite{takis,kostas}
to calculate chiral anomalies when twisted boundary conditions are present.
The starting point was the expression that incorporates all the q-loop 
contributions to the chiral anomaly in which all bosons were subjected to 
identical boundary conditions. 

A complete and meticulous examination of this problem would require steps
similar to those presented in \cite{takis,kostas} but this task is postponed
 for a future investigation. It is belevied our result to be changed  
only by a numerical factor which for the untwisted case was found to be 
$\left(\frac{-i}{2\pi}\right)^{2n}$ \footnote{In the 
untwisted case it was derived from $J=(-i)^{n/2} \prod_{\alpha =1}^{n}
\left(\Psi^{\alpha} + \Psi^{\dagger}_{\alpha} \right)$ which is the definition 
of the $\gamma_5$ analogue for the
corresponding supersymmetric quantum mechanical model, the doubling of the Majorana 
fermions, the transition element of the bosonic action, the integration over
the fermionic coherent states and the rescaling of the fermionic quantum and 
background fields.}.

\section{Appendix}

We define the generalised zeta function by the equation:
\beq
\zeta(s,a)=\sum_{n=0}^{+\infty} \frac{1}{(a+n)^s}
\feq{zeta}
  
\noi where $a$ is a constant. The contribution $I_{k_i,1}$ can be 
rewritten as:
\beq
I_{k_i,1}=\zeta(1,\alpha_{k_i})-\zeta(1,1-\alpha_{k_i}).
\feq{i1con} 

\noi The zeta function has the following asymptotic form when 
$s\rightarrow 1$:
 \beqr
\lim_{s\rightarrow 1} \left[ \zeta(s,\alpha_{k_i}) - \frac{1}{s-1}\right]
= - \frac{\Gamma^{\prime} (\alpha_{k_i})}{\Gamma (\alpha_{k_i})}.
\label{zetas}
\feqr

\noi Also from $\Gamma(z) \Gamma(1-z)=\frac{\pi}{\sin (\pi z)}$ by differentiation
with respect to z we get:
\beqr
\frac{\Gamma^{\prime}(z)}{\Gamma(z)}- \frac{\Gamma^{\prime}(1-z)}
{\Gamma(1-z)}= -\pi \cot (\pi z).
\label{difg}
\feqr

\noi Combining \rf{zetas} and \rf{difg} we arrive at \rf{i1}.

The expression \rf{baid} is easily proved with the help of:
\beq
\prod_{n=1}^{+\infty}\left[1- \left(\frac{z}{n}\right)^2 \right]=
\frac{\sin (\pi z)}{\pi z}.
\feq{exsi}

\noi One then has:
\beqr
\,\,\,&\,& \left(1\pm \frac{z}{\alpha_{k_i}} \right) \prod_{n=1}^{+\infty}
\left(1\pm \frac{z}{n+\alpha_{k_i}} \right)
\left(1\mp \frac{z}{n-\alpha_{k_i}} \right) \non \\
&=& \left(1\pm \frac{z}{\alpha_{k_i}} \right) \prod_{n=1}^{+\infty}
\left[\frac{1-\left(\frac{z\pm \alpha_{k_i}}{n} \right)^2}
{1-\left( \frac{\alpha_{k_i}}{n} \right)^2} \right] 
=\frac{\sin \pi(z\pm \alpha_{k_i})}{\sin (\pi \alpha_{k_i})}.
\label{provsi} 
\feqr

%\section{Acknowledgements}

%I would like to thank ...

\bibliographystyle{plain}

\end{document}